Research Article

# Meta-Analysis of Commercial-Scale Trials as a Means to Improve Decision-Making Processes in the Poultry Industry: A Phytogenic Feed Additive Case Study

[1]Diego A. Martinez, [2]Carol L. Ponce-de-Leon and [1]Carlos Vilchez

[1]Department of Nutrition, Universidad Nacional Agraria La Molina, Lima, Peru
[2]Independent Researcher, Lima, Peru

## Abstract

**Background and Objective:** In the current study, we sought to determine the value of a meta-analysis to improve decision-making processes related to nutrition in the poultry industry. To this end, nine commercial size experiments were conducted to test the effect of a phytogenic feed additive and three approaches were applied to the data. **Materials and Methods:** In all experiments, 1-day-old male Cobb 500 chicks were used and fed corn-soybean meal diets. Two dietary treatments were tested: T1, control diet and T2, control diet + feed additive at a 0.05% inclusion rate. The experimental units were broiler houses (7 experiments), floor pens (1 experiment) and cages (1 experiment). The response variables were final body weight, feed intake, feed conversion ratio, mortality and production efficiency. Analyses of variance of data from each and all the experiments were performed using SAS under completely randomized non-blocked or blocked designs, respectively. The meta-analyses were performed in R programming language. **Results:** No statistically significant effects were found in the evaluated variables in any of the independent experiments ($p>0.12$), nor following the application of a block design ($p>0.08$). The meta-analyses showed no statistically significant global effects in terms of final body weight ($p>0.19$), feed intake ($p>0.23$), mortality ($p>0.09$), or European Production Efficiency Factor ($p>0.08$); however, a positive global effect was found with respect to feed conversion ratio ($p<0.046$). **Conclusion:** This meta-analysis demonstrated that the phytogenic feed additive improved the efficiency of birds to convert feed to body weight (35 g less feed per 1 kg of body weight obtained). Thus, the use of meta-analyses in commercial-scale poultry trials can increase statistical power and as a result, help to detect statistical differences if they exist.

## INTRODUCTION

The feed conversion ratio (FCR) is, in addition to the cost of the feed, the most influential variable in the cost structure of poultry production[1,2] and consequently, drives the economic efficiency of poultry operations. As a result, the FCR represents an important response variable in nutrition experiments, irrespective of whether or not they are complex, as metabolism studies[3,4], or as simple as the experiments that are usually conducted to evaluate nutritional and feeding interventions. Standard nutritional experiments are frequently used to examine changes in nutrient requirements[5], use of supra-nutritional nutrient levels to modulate physiological responses, the inclusion of feed additives to optimize performance and the application of feeding strategies in broilers or layer hens[6].

The trend to produce antibiotic-free broilers is pressing the allied industry to develop technologies that help to overcome the multimodal action mechanisms of antimicrobial growth promoters[7]. An important area of research is related to the use of plant-derived products (phytogenics) to exert positive effects. Indeed, oregano (*Origanum vulgare*) represents a widely studied plant-derivative, as its essential oil and its main secondary metabolites (carvacrol and thymol)[8] have shown several biologically important activities, including antimicrobial[9,10], antioxidant[11,12], endogenous enzyme activity promoting[13,14] and prebiotic[15] properties, as well as its ability to promote intestinal mucosa structure and health[16] and prevent coccidia[17,18]. However, the overall effect of oregano essential oil on broiler performance could be challenge-dependent[19] and may vary if the chemical composition is inconstant[20].

In this regard, it becomes a complex task to perform an experiment to test these technologies, while also satisfying statistical power and meeting growing conditions similar to the industry, where natural pathogenic challenges limit the expression of the genetic potential[21]. The main reason for this is that the larger the experimental unit, the lower the statistical power, as less experimental units will be available[22]. In contrast, statistical power can be increased if more replications are made available using smaller floor pens or cages; however, the growing conditions would become less similar to the commercial ones, which would lower the challenging conditions.

One of the main limiting aspects faced by the industry and researchers is to design experiments that are sensitive enough to detect numerically small effects[3,4], such as those expected in FCR when phytogenic feed additives are tested. Usually, most of these can be economically justified with an improvement in FCR lower than 1.5%; however, the design of experiments offering such statistical sensitivity is not only a complex task[23] but is also rare. As a result, detecting these small effects becomes extremely unlikely if the study is performed under commercial conditions to test a particular technology in a real usage scenario.

In this context, meta-analysis of independent studies has been proposed as a strategy to increase statistical power[24,25]. Consequently, this is expected to support decision-making processes based on commercial-scale experiments where statistical sensitivity is insufficient, or when the expected effect is relatively low but still economically relevant. Therefore, the objective of this case study was to determine the overall effect of a phytogenic feed additive on the performance variables of broilers. In addition, we sought to compare these results with those from independent experiments included in the analysis.

## MATERIALS AND METHODS

**Experiments:** Nine independent experiments (EX1 to EX9) were performed and included a total of 622,496 broilers (Table 1). In all experiments, 1-day-old male Cobb 500 chicks were used, from 1-42 days of age. Within each experiment, birds were randomly allocated to the experimental units (EU): Whole broiler houses (7 experiments), floor pens (1 experiment), or cages (1 experiment). In experiments, reused litter based on rice husk was used as a bedding material and when cages were used, a screen was placed over the floor wiring to successfully retain the litter. In the nine experiments, corn-soybean meal-based pelleted diets that were formulated following the nutritional guidelines of the genetic line[26], were fed *ad libitum* to the birds under a four-phase feeding program (pre-starter, 0-8 day; starter, 9-18 day; grower, 19-28 day; finisher, 29-42 day) as shown in Table 2. Two dietary treatments were tested: T1, control diet and T2, control diet+ the additive at a 0.05% inclusion rate, fed continuously from 1-42 day. In all cases, the treatments were randomly assigned to the EUs. The tested phytogenic oregano-derived commercial product (blind-coded as PHE780 by LIAN Development and Service Co., Lima, Peru) provided no less than 45 g of carvacrol per kg of product.

The response variables were final body weight (BW, g bird$^{-1}$), feed intake (FI, g bird$^{-1}$), FCR (g g$^{-1}$) mortality

$$FCR = \frac{\text{Total FI per EU}}{\text{Total BW per EU}}$$

(%) and European Production Efficiency Factor (EPEF) following the calculation reported by Marcu *et al.*[27]

Table 1: Characteristics of the experiments used to test the effect of a phytogenic feed additive on the performance of broilers

| Experiment | Experimental unit | Replications per treatment | Birds per replication | Total birds |
|---|---|---|---|---|
| EX1 | Broiler house | 3 | 14,000 | 84,000 |
| EX2 | Floor pen | 5 | 40 | 400 |
| EX3 | Broiler house | 4 | 17,000 | 136,000 |
| EX4 | Broiler house | 3 | 12,000 | 72,000 |
| EX5 | Cage | 6 | 8 | 96 |
| EX6 | Broiler house | 4 | 15,000 | 120,000 |
| EX7 | Broiler house | 2 | 17,000 | 68,000 |
| EX8 | Broiler house | 3 | 16,000 | 96,000 |
| EX9 | Broiler house | 2 | 11,500 | 46,000 |

Table 2: Characteristics of the control diets used in the nine experiments conducted to determine the effect of a phytogenic feed additive on the performance of broilers[1]

| Criteria | Pre-starter (0-8 d) | Starter (9-18 d) | Grower (19-28 d) | Finisher (29-42 d) |
|---|---|---|---|---|
| **Main ingredients[2]** | | | | |
| Corn (%) | 56.80 | 59.00 | 61.70 | 64.90 |
| Soybean meal (%) | 36.50 | 34.10 | 31.20 | 27.70 |
| Soybean oil (%) | 2.51 | 2.89 | 3.34 | 3.87 |
| **Calculated nutritional content** | | | | |
| ME[3] (kcal kg$^{-1}$) | 2,975.00 | 3,028.00 | 3,090.00 | 3,165.00 |
| Crude protein (%) | 22.20 | 21.17 | 19.96 | 18.50 |
| Digestible lysine (%) | 1.25 | 1.19 | 1.11 | 1.03 |
| Non-phytic phosphorus (%) | 0.45 | 0.43 | 0.41 | 0.38 |
| Ca to non-phytic P ratio | 2.00 | 1.99 | 1.99 | 2.00 |

[1]Diets were the same for all nine experiments. [2]All phases included dicalcium phosphate, limestone, salt, synthetic amino acids (DL-methionine, L-lysine HCl, L-threonine), vitamin-mineral premix, choline chloride, mycotoxin binder, antimicrobial growth promoter (0-35 days), anticoccidial and phytase (250 FTU kg$^{-1}$ feed; partially replacing dicalcium phosphate). [3]Metabolizable energy.

$$EPEF = \frac{BW, kg \times (100 - mortality, \%) \times 100}{FCR \times age, d}$$

In broiler houses, the BW was obtained by weighing 10 sub-samples of 50 birds, each one in different locations within the house and the FI was calculated assuming that all of the feed provided was eaten. In the floor pens and cages, the BW was obtained by weighing all the birds and the FI was calculated as the actual net amount of feed eaten.

**Analyses of variance:** Data were first analyzed independently by experiment under completely randomized designs and thereafter, data were combined and analyzed under a completely randomized block design, considering the experiment itself as the blocking factor[28]. Normality of the data was determined using the Shapiro-Wilk test[29] and the existence of outliers was determined by Grubbs test[30]. The response variables with non-normal distributions were analyzed with Kruskal-Wallis test[31]. In all cases, results were considered statistically significant when $p \leq 0.05$.

The additive linear model for the analysis of each independent experiment was $Y_{ij} = \mu + \tau_i + \epsilon_{ij}$, where $Y_{ij}$ is the observed value in the i-th treatment (i: 1,...t) and j-th replication (j: 1,...r); $\mu$ is the effect of the general mean; $\tau_i$ is the effect of the i-th treatment; $\epsilon_{ij}$ is the effect of the experimental error in the i-th treatment and j-th replication; t is the number of treatments; r is the number of replications in the i-th treatment; being that $\epsilon_{ij} \sim N(\mu,\sigma^2)$ and independently, where N denotes the normal distribution among replications and $\sigma^2$ is the variance among the experimental error of the different EU.

In contrast, the additive linear model for the analysis of variance of the whole data was $Y_{ijk} = \mu + \tau_i + \beta_j + \epsilon_{ijk}$, where $Y_{ijk}$ is the observed value in the i-th treatment (i: 1,...t), j-th block (j: 1,...p) and k-th replication (k: 1,...r); $\mu$ is the effect of the general mean; $\tau_i$ is the effect of the i-th treatment; $\beta_j$ is the effect of the j-th block; $\epsilon_{ijk}$ is the effect of the experimental error in the i-th treatment, j-th block and k-th replication; t is the number of treatments; p is the number of blocks; r is the number of replications in the i-th treatment; being that $\epsilon_{ijk} \sim N(\mu,\sigma^2)$ and independently, where N denotes the normal distribution among replications and $\sigma^2$ is the variance among the experimental error of the different EU.

**Meta-analyses:** Independent meta-analyses were performed for each single response variable to determine the overall effect size, its 95% confidence interval (CI$_{95\%}$) and its probability with Wald test[32] and the existence of heterogeneity using a random-effects model with Cochran test[33] (Q statistic) and its corresponding probability with chi-square test[34]. In all cases, results were considered statistically significant when $p \leq 0.05$.

The heterogeneity was determined considering the following linear additive model: $y_i = \mu + u_i + e_i$, where yi is the observed effect size in the i-th experiment (i: 1,...k) (and also, $y_i = \theta_i + e_i$, where $\theta_i$ is the unknown true effect in the i-th experiment; $e_i$ is the intra-experimental sampling error in the i-th experiment); $u_i$ is the inter-experimental deviation regarding the overall effect size in the i-th experiment; $e_i$ is the intra-experimental sampling error in the i-th study; k is the number of experiments; N denotes the normal distribution of the random inter-experimental deviation (u) and the intra-experimental sampling error (e); being that $u_i \sim N(0, \tau^2)$ y $e_i \sim N(0, v_i)$ and both independently, where $\tau^2$ indicates the heterogeneity (variability among the true effects in the different experiments) and $v_i$ is the approximately known sampling variance of the estimated effect size in the i-th experiment.

To adjust the model, a weighted least square method was applied, implying that the adjusted model provides an estimate of $\bar{\theta}_w = \sum w_i \theta_i / \sum w_i$, where is the true weighted average effect size; $w_i$ is the weighing factor considered, $\theta_i$ is the true effect size in the i-th experiment; that is, is the weighted average of the true effects ($\theta_i$) in the set of k studies, with weights equal to the inverse of the corresponding variances ($w_i = 1/v_i$).

In addition, the goodness of fit of model residues were evaluated with the Shapiro-Wilk test (normal if p>0.05). In cases where the residues were non-normally distributed, the data were analyzed again to determine the probability associated to the global effect size but this time, with applying a permutation test with 10,000 iterations.

Finally, the presence of bias within the data of each response variable was evaluated through the Egger regression test to determine the asymmetry of the distribution of the data, based on both the effect sizes and the precision of each experiment. Trim and Fill analysis was then performed to estimate the effect size values that would compensate distribution imbalances, if they existed, and if so, their magnitude and influence on the overall effect size were determined. As a result, each variable eventually had two sets of effect sizes: dO, being the set of effect sizes calculated from the experiments and dA, being the set of effect sizes that also included the values estimated through the Trim and Fill analysis. Thereafter, the bias was considered relevant if the Egger test was significant (p≤0.05) and if the $CI_{95\%}$ of the overall effect sizes, calculated with both the adjusted data (dA) and with the original data (dO), were not overlapped.

**Software and informatics resources:** Grubbs test for the detection of outliers was performed with GraphPad Prism 7 software[35]. Kruskal-Wallis tests and variance analyses were performed in SAS 9.4 using NPAR1WAY with Wilcoxon restriction and GLM procedures, respectively[36]. The goodness of fit to the normal distribution and meta-analyses routines were performed with *stats* and *Metafor* 2.0-0[37] packages in R 3.5.2 version programming language[38] using RStudio 1.1.456 as an interface[39].

## RESULTS

**Analyses of variance:** No outliers were detected in the data from each independent experiment; however, the Shapiro-Wilk goodness of fit test showed non-normally distributed mortality values; therefore, the data of this variable were analyzed with the Kruskal-Wallis test. The results found in each of the nine experiments and in the combined analysis are shown in Table 3. The highest percentage differences between treatments in BW, FCR and EPEF were +5.28, -4.50 and +6.65% in experiments EX6, EX5 and EX1, respectively; however, even these differences were not statistically significant (p>0.12). Similarly, the combined analysis of the nine experiments under a completely randomised block design showed no statistically significant effects on any of the tested variables (p>0.08).

**Meta-analyses results:** Table 4 shows the meta-analyses results. Test for the goodness of fit of model residuals found mortality values being non-normally distributed; therefore, the overall effect size p-value for this variable was recalculated by applying a permutation test. BW (Fig. 1), FI (Fig. 2), mortality (Fig. 3) and EPEF (Fig. 4) showed no significant (p>0.05) overall effect sizes and had $CI_{95\%}$ with limit values with opposite mathematical signs (positive, negative). A statistically significant (p<0.05) overall effect size was found in FCR (Fig. 5), with a $CI_{95\%}$ with negative limit values. The Trim and Fill tests determined and estimated possibly missing BW, FI and mortality values; however, the $CI_{95\%}$ of the adjusted overall effect sizes for all these variables, were overlapped with the $CI_{95\%}$ calculated with the original data; therefore, if biases existed, they were not considered to be relevant. In addition, Egger tests did not detect statistically significant bias (p>0.50) and no statistically significant heterogeneity was found among experiments in any of the tested variables (p>0.23).

## DISCUSSION

This study investigated the effect of a phytogenic feed additive on the performance of broilers. We sought to explore three different approaches to analyze the data from nine experiments to increase the likelihood of finding statistically significant effects, if they existed. The aim of this study was to determine a suitable method to improve decision-making processes related to nutrition and feeding strategies in the poultry industry.

Table 3: Effect of a phytogenic feed additive on the performance of 42-day-old broilers[1]

| Treatments[2] | EX1 | EX2 | EX3 | EX4 | EX5 | EX6 | EX7 | EX8 | EX9 | All |
|---|---|---|---|---|---|---|---|---|---|---|
| **Final body weight (BW) (kg bird$^{-1}$)** | | | | | | | | | | |
| T1 | 2.632 | 3.132 | 2.878 | 2.821 | 2.924 | 2.880 | 2.700 | 2.897 | 2.859 | 2.888 |
| T2 | 2.708 | 3.221 | 2.943 | 2.751 | 3.038 | 3.032 | 2.665 | 2.939 | 2.825 | 2.950 |
| Difference (%) | 2.8 | 2.8 | 2.2 | -2.4 | 3.8 | 5.2 | -1.2 | 1.4 | -1.1 | 2.1 |
| p-value | 0.658 | 0.583 | 0.728 | 0.653 | 0.299 | 0.129 | 0.775 | 0.634 | 0.771 | 0.165 |
| SEM[3] | 0.194 | 0.248 | 0.251 | 0.176 | 0.179 | 0.122 | 0.106 | 0.102 | 0.102 | 0.178 |
| **Feed intake (FI) (kg bird$^{-1}$)** | | | | | | | | | | |
| T1 | 4.786 | 5.511 | 4.966 | 4.974 | 5.067 | 5.195 | 4.891 | 5.047 | 4.928 | 5.083 |
| T2 | 4.767 | 5.769 | 4.906 | 4.740 | 5.037 | 5.378 | 4.859 | 5.034 | 4.728 | 5.094 |
| Difference (%) | -0.4 | 4.6 | -1.2 | -4.7 | -0.5 | 3.5 | -0.6 | -0.2 | -4.0 | 0.2 |
| p-value | 0.966 | 0.473 | 0.772 | 0.125 | 0.878 | 0.258 | 0.612 | 0.919 | 0.301 | 0.898 |
| SEM[3] | 0.516 | 0.541 | 0.281 | 0.148 | 0.334 | 0.207 | 0.054 | 0.139 | 0.144 | 0.328 |
| **Feed conversion ratio (FCR)[4] (g g$^{-1}$)** | | | | | | | | | | |
| T1 | 1.821 | 1.765 | 1.731 | 1.766 | 1.736 | 1.807 | 1.814 | 1.744 | 1.727 | 1.765 |
| T2 | 1.754 | 1.787 | 1.669 | 1.726 | 1.658 | 1.775 | 1.824 | 1.715 | 1.674 | 1.726 |
| Difference (%) | -3.6 | 1.2 | -3.5 | -2.2 | -4.5 | -1.7 | 0.5 | -1.6 | -3.0 | -2.2 |
| p-value | 0.481 | 0.772 | 0.252 | 0.559 | 0.151 | 0.647 | 0.867 | 0.734 | 0.671 | 0.085 |
| SEM[3] | 0.107 | 0.114 | 0.068 | 0.076 | 0.087 | 0.093 | 0.053 | 0.097 | 0.108 | 0.087 |
| **Mortality[5] (%)** | | | | | | | | | | |
| T1 | 3.923 | 2.000 | 4.280 | 4.517 | 6.250 | 4.650 | 4.235 | 3.817 | 3.960 | 4.262 |
| T2 | 3.727 | 2.500 | 4.298 | 4.147 | 8.333 | 4.345 | 3.590 | 3.600 | 4.110 | 4.590 |
| Difference (%) | -5.0 | 25.0 | 0.4 | -8.1 | 33.3 | -6.5 | -15.2 | -5.6 | 3.7 | 7.7 |
| P-value | 0.513 | 0.650 | 0.773 | 0.513 | 0.575 | 0.309 | 0.439 | 0.376 | 1.000 | 0.904 |
| SEM[3] | 0.683 | 1.936 | 0.701 | 0.564 | 6.654 | 0.552 | 0.731 | 0.503 | 0.497 | 3.031 |
| **European production efficiency factor (EPEF)[6]** | | | | | | | | | | |
| T1 | 331.7 | 416.1 | 380.7 | 364.1 | 378.8 | 363.7 | 340.1 | 381.5 | 380.6 | 375.1 |
| T2 | 353.8 | 419.3 | 402.7 | 365.2 | 399.8 | 389.3 | 335.5 | 394.8 | 385.6 | 389.0 |
| Difference (%) | 6.6 | 0.7 | 5.8 | 0.3 | 5.5 | 7.0 | -1.3 | 3.5 | 1.3 | 3.7 |
| p-value | 0.348 | 0.912 | 0.539 | 0.972 | 0.462 | 0.262 | 0.866 | 0.658 | 0.905 | 0.145 |
| SEM[3] | 25.4 | 43.6 | 47.9 | 39.0 | 47.5 | 29.1 | 23.9 | 34.3 | 37.1 | 37.4 |

[1]EX1 to EX9: Each of the nine conducted experiments (EX). Values on columns EX1 to EX9 correspond to the average of 3, 5, 4, 3, 6, 4, 2, 3 and 2 replications, respectively, in which each experiment is treated as an independent completely randomized design. Values in the last column ("All") correspond to the average of all the data, that are treated as a completely randomized block design, with the experiment considered as the blocking factor. [2]T1: Control diet, T2: Control diet +phytogenic feed additive at a 0.05% inclusion rate. [3]SEM: Standard error of the mean. [4] $FCR = \frac{\text{Total FI per experimental unit}}{\text{Total BW per experimental unit}}$. [5]In all cases, mortality data were analyzed with the non-parametric Kruskal-Wallis test. [6] $EPEF = \frac{BW, kg \times (100 - \text{mortality}, \%) \times 100}{FCR \times \text{age, day}}$

Table 4: Meta-analyses of the effect of a phytogenic feed additive on performance of 42-day-old broilers (nine experiments)

| Criteria | Response variables[1] | | | | |
|---|---|---|---|---|---|
| | BW (kg bird$^{-1}$) | FI (kg bird$^{-1}$) | FCR | Mortality (%) | EPEF |
| **Effect size of the phytogenic feed additive, calculated with the original data** | | | | | |
| Effect size | +0.0393 | -0.0540 | -0.0346 | -0.1818 | +12.3836 |
| CI$_{95\%}$[2] | -0.0198 to +0.0983 | -0.1432 to +0.0352 | -0.0686 to -0.0006 | -0.4655 to +0.1019 | -1.7088 to +26.4761 |
| Effect size p-value[3] | 0.1928 | 0.2355 | 0.0460 | 0.0938* | 0.0850 |
| **Goodness of fit of model residuals to normal distribution** | | | | | |
| p-value[4] | 0.5850 | 0.2996 | 0.4190 | 0.0082 | 0.2377 |
| **Bias** | | | | | |
| Possible missing values | 3 | 2 | 0 | 1 | 0 |
| Adjusted effect size[5] | +0.0011 | -0.0902 | -0.0346 | -0.1857 | +12.3836 |
| Adjusted CI$_{95\%}$[5] | -0.0644 to +0.0622 | -0.1917 to +0.0112 | -0.0686 to -0.0006 | -0.4691 to +0.0978 | -1.7088 to +26.4761 |
| Bias p-value[3] | 0.7885 | 0.5079 | 0.7836 | 0.5026 | 0.9154 |
| **Heterogeneity among experiments** | | | | | |
| P-value[3] | 0.6483 | 0.2344 | 0.8746 | 0.9272 | 0.9611 |

[1]BW: Final body weight (42 day), FI: Feed intake, FCR: Feed conversion ratio, EPEF: European Production Efficiency Factor. [2]CI$_{95\%}$: Confidence interval at 95%. [3]Overall effect size, bias, or heterogeneity are statistically significant if p≤0.05. [4]Non-normal distribution if p≤0.05. [5]Calculated including the predicted possibly missing values estimated though the Trim and Fill test. *Probability estimated through the permutation test, as the model residuals were not normally distributed.

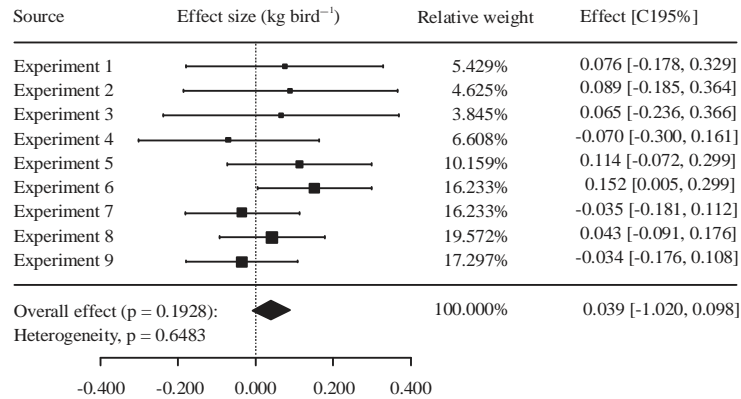

Fig. 1: Forest plot of the effects of a phytogenic feed additive on the body weight of 42-day-old broilers (nine experiments)

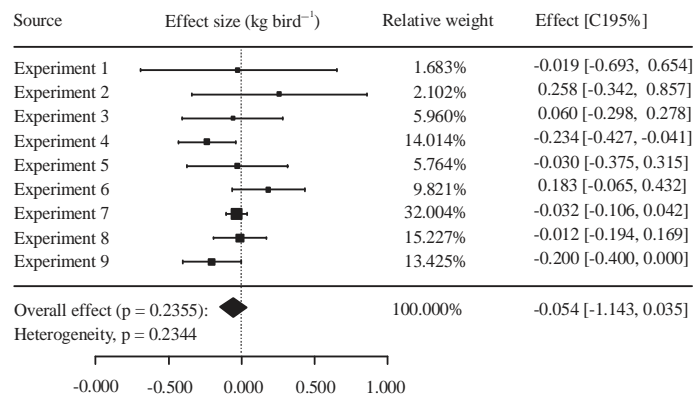

Fig. 2: Forest plot of the effects of a phytogenic feed additive on the feed intake of 42-day-old broilers (nine experiments)

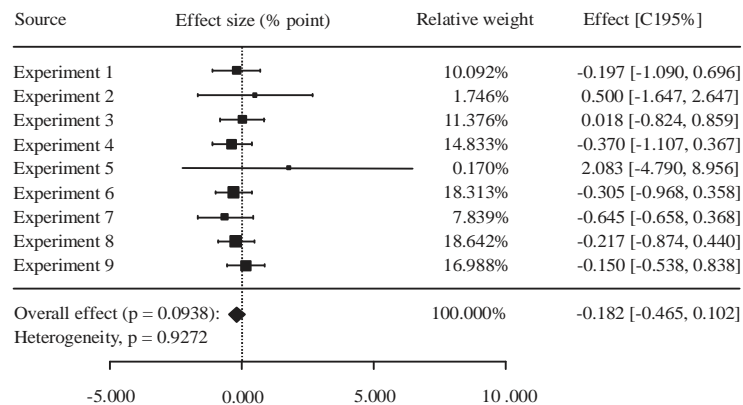

Fig. 3: Forest plot of the effects of a phytogenic feed additive on the mortality of 42-day-old broilers (nine experiments)

The results showed that neither analyzing the data from the different experiments independently under completely randomised designs, nor combining all the data under a block design, led to statistically significant effects in any of the tested variables. The lack of sensitivity to detect differences as big as +5.28, -4.50 and +6.65% in BW, FCR and EPEF, respectively, was influenced by the low number of replications used in the experiments[21]. However, this is the usual scenario faced by the industry when evaluating nutrition or feeding strategies under actual commercial-scale conditions[40]. Under such situations, there are three main consequences: (1) Companies take positive decisions but without the desired confidence and, consequently, become short-lived, (2) decision-making processes become complex and longer; or (3) no decision is taken, status quo is maintained and the opportunity to improve results may be lost. In addition, it has been reported that when it is more difficult for a person to make decisions based on rigorous reasoning, it ultimately leads to a more intuitive and heuristic thinking process due to decision fatigue; consequently, less judicious decisions are taken[41].

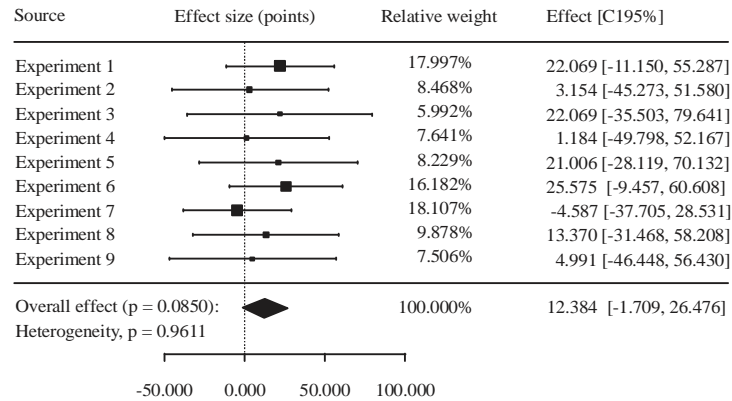

Fig. 4: Forest plot of the effects of a phytogenic feed additive on the European production efficiency factor of 42-day-old broilers (nine experiments)

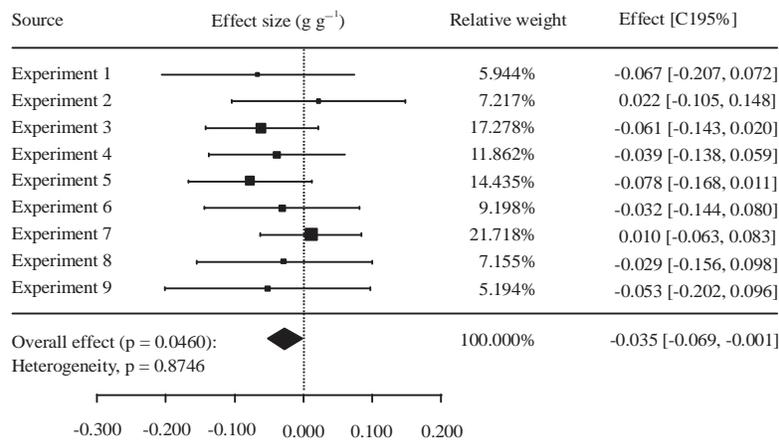

Fig. 5: Forest plot of the effects of a phytogenic feed additive on the feed conversion ratio of 42-day-old broilers (nine experiments)

Although none of the independent experiments showed significant effects on the studied variables (p>0.05), this should not be interpreted as that the evaluated product does not produce an effect on these response variables. Instead, this may be explained by the fact that in hypothesis testing, the null hypothesis (that both means are equal) can only be rejected and not proved[42,43]. In this regard, under the Neyman-Pearson dichotomous approach, a p-value greater than the pre-established $\alpha$ level of significance in a hypothesis test of the difference of two means determines that the null hypothesis must be exhaustively accepted as true. However, the Fischer approach considers the p-value as a continuous measure of the strength of evidence[44] and states that the absence of a significant effect could only indicate that, if such an effect exists, it is not sufficiently large to be detected by an experiment of the size used[45].

Although, the meta-analyses did not detect effects on BW, FI, mortality, or EPEF, we demonstrated an improvement in FCR that was due to the feed additive tested, in that the supplemented birds converted feed to body weight more efficiently (35 g less feed per kg body weight obtained). No significant heterogeneity was detected among experiments (p>0.87), indicating that the effect of the feed additive on the FCR was not inconsistent across the nine experiments. In addition, the CI of the effect size in FCR (0.0006 to 0.0686 less FCR points) indicates that, regardless of the accuracy of the estimation of the effect, the real effect of the phytogenic on feed efficiency is positive[43].

The effect of the tested phytogenic feed additive found in FCR agrees with previous reports about the effect of oregano essential oil on the FCR of broilers[17,46-49]. This finding is consistent with the antimicrobial[9,10], antioxidant[11-13], endogenous enzyme activity promoting[10,14], prebiotic[15], anticoccidial[18] and gut mucosa promoting effects[16] of oregano essential oil that have been previously shown. Besides, previous studies have reported positive effects of oregano essential oil on intestinal mucosa structure, nutrient absorption capacity, bone mineralization and overall performance[50]. It has been reported that the effect of oregano essential oil on broiler performance could be challenge-dependent[19]; however, in the current study, the experiments were conducted under commercial conditions, unavoidably implying certain intestinal challenges, since reused litter material was used in all experiments[51-53].

In the present study, when combining the data from all the experiments under a completely randomized block design, the statistical power for FCR increased and, therefore, the p-value (p = 0.085) was lower in comparison to that observed in individual experiments (p-values: 0.151-0.867). However, the p-value was not only not considered significant but also was 85% higher than that obtained for the overall effect on FCR through meta-analysis (p = 0.046).

Thus, in the present analysis, we demonstrate how meta-analyses of the results obtained in different experiments favour the probability of detecting an effect, when it exists, that may not be evident in independent experiments. In this regard, although the meta-analyses carried out using random effects models do not guarantee that the inclusion of additional studies increases the statistical power of the analysis, in general, it does increase the statistical power in comparison to the independent studies[24,25]. This is particularly useful when the critical response variable in an experiment is FCR, as usually a small percentage effect, even less than 2%[54], is sufficient for the poultry producer to justify making a favourable decision regarding the nutritional benefit of the feeding strategy tested. In addition, significant effect sizes obtained by meta-analysis also allows the nutritionist to make a cost-sensitivity analysis[55,56]. Previous meta-analyses have detected small percentage effects on FCR in broilers[54], layer hens[57] and pigs[58]; however, to the best of our knowledge, a meta-analysis approach has not yet been reported for analysing commercial size trials with a low number of replications to help improve statistical sensitivity.

Experiment standardization is a common strategy to increase the sensitivity of the test; however, this also reduces the reproducibility of the results[59]. In this regard, meta-analysis of commercial-scale experiments not only allows the sensitivity of the analysis to be increased[24] but also preserves the reproducibility of the results, as they are performed in conditions less homogeneous than those of a highly controlled research facility. Therefore, the higher the systematic variation, the greater the reproducibility of the experiment[59]. Finally, in poultry nutrition research, statistical sensitivity and growing conditions similar to the industry are commonly opposite objectives, as the more sensitive a design is, the more replications it takes and the smaller they become[21]; however, a meta-analysis can go some way to help solve this dichotomy.

## CONCLUSION

In the present study, we tested a phytogenic feed additive, based on oregano essential oil, providing no less than 45 g carvacrol per kg of product and fed at an inclusion rate of 0.05%, continuously from 1-42 day. Based on the observed results, it can be concluded that the tested product improved the FCR of broilers under commercial-scale conditions, in that it increased the efficiency of converting feed into BW (35 g less feed per 1 kg of BW obtained). In addition, the analysis of the nine conducted experiments using a meta-analysis approach improved the statistical power to a greater magnitude than that observed by applying a block design. Moreover, the meta-analysis was sensitive enough to detect a statistical significance that, otherwise, would have remained undetected.

## SIGNIFICANCE STATEMENT

This study demonstrated that meta-analysis is a useful technique to improve statistical power and to help find statistically significant differences, if they exist, when testing nutritional interventions under commercial conditions. We postulate that the use of meta-analysis in the poultry industry would help industry nutritionists and researchers to establish a more efficient but still simple, system to evaluate nutrition interventions, including feed additives and consequently, provide a means to facilitate and objectivize decision-making processes.

**ACKNOWLEDGMENTS**

The authors would like to thank LIAN Development and Service Co., Lima, Peru, for partial financial support of this research.

Int. J. Poult. Sci., 19 (11): 513-523, 202025. Borenstein, M., L.V. Hedges, J.P.T. Higgins and H.R. Rothstein, 2009. Power Analysis for Meta Analysis. In: Introduction to Meta Analysis. Borenstein, M., L.V. Hedges, J.P.T. Higgins and H.R. Rothstein (Eds.). John Wiley & Sons, Ltd Hoboken, New Jersey pp: 257-276.
26. Cobb-Vantress, 2015. Broiler performance and nutrition supplement. July 2015. http://www.cobb-vantress.com/docs/default-source/cobb-500-guides/Cobb500_Broiler_Performance_And_Nutrition_Supplement.pdf.
27. Marcu, A., I. Vacaru-Opri, G. Dumitrescu, L.P. Ciochină and A. Marcu et al., 2013. The influence of genetics on economic efficiency of broiler chickens growth. Anim. Sci. Biotechnol., 46: 339-346.
28. Petschenka, G., C. Pick, V. Wagschal and S. Dobler, 2013. Functional evidence for physiological mechanisms to circumvent neurotoxicity of cardenolides in an adapted and a non-adapted hawk-moth species. Proc. R. Soc. B, Vol. 280 10.1098/rspb.2012.3089
29. Adler, J., 2009. R in a Nutshell: A Desktop Quick Reference. 1st Edn., O'Reilly Media Sebastopol, California.
30. Grubbs, F.E., 1969. Procedures for detecting outlying observations in samples. Technometrics, 11: 1-21.
31. McDonald, J.H., 2009. Handbook of Biological Statistics. 2nd Edn., Sparky House Publishing, Baltimore, Maryland.
32. Noortgate, V.d., W. Onghena and Patrick, 2003. Multilevel meta-analysis: A comparison with traditional meta-analytical procedures. Educ. Psychol. Meas., 63: 765-790.
33. Hedges, L.V. and I. Olkin, 1985. Parametric Estimation of Effect Size from a Series of Experiments. In: Statistical Methods for Meta-Analysis. Hedges, L.V. and I. Olkin, Academic Press Inc., United States of America pp: 107-145.
34. Bougouin, A., J.A.D.R.N. Appuhamy, E. Kebreab, J. Dijkstra, R.P. Kwakkel and J. France, 2014. Effects of phytase supplementation on phosphorus retention in broilers and layers: A meta-analysis. Poult. Sci., 93: 1981-1992.
35. Motulsky, H.J., 2007. Graphpad prism 5.00 Statistics Guide. GraphPad Software, inc. https://cdn.graphpad.com/faq/2/file/Prism_v5_Statistics_Guide.pdf
36. SAS, 2017. Base SAS® 9.4 procedures guide. 7th Edn., SAS Institute, Inc., North Carolina, USA Pages: 175.
37. Viechtbauer, W., 2010. Conducting meta-analyses in R with the metafor package. J. Stat. Software, Vol. 36, No. 3 10.18637/jss.v036.i03
38. R Core Team., 2018. R: A language and environment for statistical computing. R Foundation for Statistical Computing, Vienna, Austria.
39. RStudio, 2016. RStudio: integrated development for R. RStudio Inc.,. https://rstudio.com
40. Dawkins, M.S., 2012. Commercial scale research and assessment of poultry welfare. Br. Poult. Sci., 53: 1-6.
41. Hirshleifer, D., Y. Levi, B. Lourie and S.H. Teoha, 2019. Decision fatigue and heuristic analyst forecasts. J. Financial Econ., 133: 83-98.
42. Nicholls, N., 2001. The Insignificance of Significance Testing. Bull. Amer. Meteor. Soc., 82: 981-986.
43. Blackwelder, W.C., 1982. "Proving the null hypothesis" in clinical trials. Controlled Clin. Trials, 3: 345-353.
44. Dragicevic, P., 2016. Fair Statistical Communication in HCI. In: Modern Statistical Methods for HCI. Robertson, J. and M. Kaptein (Eds.). Springer, Cham Switzerland pp: 291-330.
45. Chris, O. and W.H. Freeman, 2002. The Lady Tasting Tea: How Statistics Revolutionized Science in the Twentieth Century. American Statistical Association United States of America Pages: 650.
46. Roofchaee, A., M. Irani, M.A. Ebrahimzadeh and M.R. Akbari, 2011. Effect of dietary oregano (*Origanum vulgare* L.) essential oil on growth performance, cecal microflora and serum antioxidant activity of broiler chickens. Afr. J. Biotechnol., 10: 6177-6183.
47. Cho, J.H., H.J. Kim and I.H. Kim, 2014. Effects of phytogenic feed additive on growth performance, digestibility, blood metabolites, intestinal microbiota, meat color and relative organ weight after oral challenge with *Clostridium perfringens* in broilers. Livest. Sci., 160: 82-88.
48. Ghazi, S., T. Amjadian and S. Norouzi, 2014. Single and combined effects of vitamin C and oregano essential oil in diet, on growth performance and blood parameters of broiler chicks reared under heat stress condition. Int. J. Biometeorol., 58: 741-752.
49. Peng, Q.Y., J.D. Li, Z. Li, Z.Y. Duan and Y.P. Wu, 2016. Effects of dietary supplementation with oregano essential oil on growth performance, carcass traits and jejunal morphology in broiler chickens. Anim. Feed Sci. Technol., 214: 148-153.
50. Martinez, D.A., 2012. Evaluación de un producto a base de aceite esencial de orégano sobre la integridad intestinal, la capacidad de absorción de nutrientes y el comportamiento productivo de pollos de carne. Master Thesis, Universidad Nacional Agraria La Molina.
51. Bregendahl, K., D.U. Ahn, D.W. Trampel and J.M. Campbell, 2005. Effects of dietary spray-dried bovine plasma protein on broiler growth performance and breast-meat yield. J. Applied Poult. Res., 14: 560-568.
52. Barker, K.J., J.L. Purswell, J.D. Davis, H.M. Parker, M.T. Kidd, C.D. McDaniel and A.S. Kiess, 2010. Distribution of bacteria at different poultry litter depths[1]. Int. J. Poult. Sci., 9: 10-13.
53. O'Reilly, E.L., R.J. Burchmore, N.H. Sparks and P.D. Eckersall, 2016. The effect of microbial challenge on the intestinal proteome of broiler chickens. Proteome Sci., Vol. 15, No. 10 10.1186/s12953-017-0118-0
54. Gaibor, J.R.Q., R. Torres, J. Yupanqui, D.M. Patiño-Patroni and C.V. Perales, 2019. Suplementación alimenticia de glutamina sobre el desempeño productivo en pollos de engorde. Siembra, 6: 15-24.
55. Campbell, M.K. and D.J. Torgerson, 1999. Bootstrapping: Estimating confidence intervals for cost-effectiveness ratios. Int. J. Med., 92: 177-182.
522